\documentclass[11pt]{article}

\usepackage{arxiv}

\usepackage[utf8]{inputenc} 
\usepackage[T1]{fontenc}    
\usepackage{hyperref}       
\usepackage{breakurl}
\usepackage{adjustbox}
\usepackage{url}            
\usepackage{booktabs}       
\usepackage{amsfonts}       
\usepackage{amssymb}
\usepackage{nicefrac}       
\usepackage{microtype}      
\usepackage{graphicx}
\usepackage{csquotes}
\usepackage{doi}
\usepackage{amsmath}
\usepackage{cleveref}       
\usepackage{enumitem}
\usepackage{xcolor}

\title{Reconciling Open Interest with Traded Volume in Perpetual Swaps}

\author{\href{https://orcid.org/0000-0002-5521-2196}{\hspace{1mm}Ioannis ~Giagkiozis} \\
	Chrysor Trading\\
	Abu Dhabi, UAE \\
	\texttt{ioannis@chrysor-trading.com} \\
	\And
	\href{https://orcid.org/0000-0001-7166-5016}{\hspace{1mm}Emilio ~Said} \\
	Abu Dhabi Investment Authority\\
	Strategy and Planning Department\\
	Quantitative Research and Development \\
	Abu Dhabi, UAE \\
	\texttt{emilio.said@adia.ae}
}

\renewcommand{\headeright}{}
\renewcommand{\undertitle}{\textit{(Published in Ledger, Volume 9 (2024), 1-15)}}
\renewcommand{\shorttitle}{Reconciling Open Interest with Traded Volume}

\hypersetup{
pdftitle={Reconciling open interest with traded volume in perpetual swaps},
pdfsubject={q-bio.NC, q-bio.QM},
pdfauthor={Ioannis ~Giagkiozis, Emilio ~Said},
pdfkeywords={crypto,btc,bitcoin,derivatives,exchange,open interest,funding rates},
}

\DeclareRobustCommand{\btcusdt}{BTC\_USDT\_P}
\DeclareRobustCommand{\btcusd}{BTC\_USD\_IP}
\DeclareRobustCommand{\bybit}{ByBit}
\DeclareRobustCommand{\bitmex}{BitMEX}
\DeclareRobustCommand{\eqref}[1]{Eqn.~(\ref{#1})}
\DeclareRobustCommand{\tblref}[1]{Table~\ref{#1}}

\def\bitcoin{%
  \leavevmode
  \vtop{\offinterlineskip 
    \setbox0=\hbox{B}%
    \setbox2=\hbox to\wd0{\hfil\hskip-.03em
    \vrule height .3ex width .15ex\hskip .08em
    \vrule height .3ex width .15ex\hfil}
    \vbox{\copy2\box0}\box2}}

\begin{document}
\maketitle

\begin{abstract}
Perpetual swaps are derivative contracts that allow traders to speculate on, or hedge, the price movements of cryptocurrencies. Unlike futures contracts, perpetual swaps have no settlement or expiration in the traditional sense. The \textit{funding rate} acts as the mechanism that tethers the perpetual swap to its underlying with the help of arbitrageurs. Open interest, in the context of perpetual swaps and derivative contracts in general, refers to the total number of outstanding contracts at a given point in time. It is a critical metric in derivatives markets as it can provide insight into market activity, sentiment and overall liquidity. It also provides a way to estimate a lower bound on the collateral required for every cryptocurrency market on an exchange. This number, cumulated across all markets on the exchange in combination with proof of reserves, can be used to gauge whether the exchange in question operates with unsustainable levels of leverage, which could have solvency implications. We find that open interest in Bitcoin perpetual swaps is systematically misquoted by some of the largest derivatives exchanges; however, the degree varies, with some exchanges reporting open interest that is wholly implausible to others that seem to be delaying messages of forced trades, \textit{i.e.}, liquidations. We identify these incongruities by analyzing tick-by-tick data for two time periods in $2023$ by connecting directly to seven of the most liquid cryptocurrency derivatives exchanges.
\end{abstract}


\section{Introduction}
\label{sec:introduction}

Perpetual swaps were introduced by BitMEX in $2016$ \cite{hayesAnnouncingLaunchPerpetual2016}. They are futures contracts with no expiry. These contracts allow for high leverage with most cryptocurrency exchanges offering leverage in the range of $100$x--$125$x and some recent platforms allowing up-to $1000$x(!) leverage\footnote{Reference intentionally omitted as trading with such high leverage, especially in volatile markets like cryptocurrencies, is ill-advised. $1000\mathbf{x}$ leverage implies, in the best case, that the liquidation price is $10$ basis points away from the entry price and more realistically $5$ basis points considering fees}. Interestingly a reduction in allowed leverage, in retrospect, can be a sign that an exchange is in distress \cite{reynoldsFTXCutsLeverage2021}. These contracts are designed to track an underlying exchange rate, \textit{e.g.} BTC / USD, such that speculators can gain exposure to that underlying while holding a collateral of their choice (usually \href{https://tether.to/en/}{USDT}). The first perpetual swap was what is commonly referred to today as \textit{inverse perpetual}. In inverse perpetual contracts, profits and losses as well as margin are paid in the base asset, \textit{e.g.}, BTC for the BTC/USD inverse perpetual, while the price of the contract is quoted in units of the quote asset. Except for their use as an instrument for speculation, this kind of perpetual swap was originally also used as a tool to hedge exposure to the underlying. This can be achieved by opening a short position in the contract with $1$x leverage. As the inverse perpetual is \textit{coin} margined, every change in the price of the underlying is offset by the short position in the inverse perpetual which results in a stable equity curve when denominated in units of the quote asset. Before \href{https://tether.to/en/}{USDT} was accepted as the \textit{de facto} stablecoin in cryptocurrency trading, this was the primary mechanism for traders to hedge their Bitcoin exposure. 

As stablecoins gained in popularity, inverse perpetuals ceded market dominance to the \textit{linear perpetual} swap. These pay profits and losses in the quote asset and are margined by the same, which is most often some stablecoin for the US dollar. This is not unexpected, as for most people their numéraire is some form of FIAT currency, predominantly USD, especially when it comes to financial instruments. However, the benefits of the linear perpetual are mostly limited to their use as an instrument for speculation, given that these contracts are margined by the quote asset. This margining implies the existence of a liquidation price which necessitates active monitoring and rebalancing if used as a hedge. In addition, most exchanges have a limit on the size of the position that can be opened on linear perpetuals, which reduces the capacity of a potential hedge.

Originally, the mechanism used to tether perpetual swaps to the underlying was the interest rate differential for USD and Bitcoin on the Bitfinex lending market \cite{hayesAnnouncingLaunchPerpetual2016}. Namely, if it was relatively more expensive to borrow USD than Bitcoin, then long positions paid short positions the difference, in lending rates multiplied by the individual position size, and shorts paid long positions if the borrow rate of Bitcoin was higher than USD. These cashflows were automatically exchanged every 8 hours among market participants who held positions at the 8-hour mark, with no fees charged by the exchange. At the time, cryptocurrency markets were not as strongly coupled as they are presently, which led to frequent dislocations between the price of the contract and the underlying. Today, most exchanges use some variation of what is referred to as the \textit{funding rate}. The funding rate is computed as a function of the perpetual swap price $F_t$, and an index price which is a weighted average of the SPOT market of the underlying, $S_t$, over a number of exchanges. The details vary from exchange to exchange; however, this function is constructed in such a way as to incentivize short positions by means of funding payments when $F_t > S_t$ and long positions when $F_t < S_t$. Price dislocations with this mechanism can and do happen, but they are much less frequent and smaller in magnitude \cite{heFundamentalsPerpetualFutures2023}.

Centralized cryptocurrency exchanges offer their market data feeds freely. The majority give access to level 2 data even to anonymous clients: with the appropriate software anyone can collect the data reported by these exchanges in real-time. Some exchanges even offer historical archival services of tick level data (see the Appendix). This is in stark contrast to traditional exchanges, \textit{e.g.} the NASDAQ and NYSE, whose historical and real-time tick-by-tick market data and feeds are behind exorbitant paywalls. From that perspective, cryptocurrency exchanges have introduced a much higher standard of transparency and inclusivity---something that we believe ought to be acknowledged, especially in an environment where narratives tend to be highly polarized and one-sided. That said, the record of cryptocurrency exchanges is far from spotless. There have been numerous reports alleging fake volume \cite{InvestigationLegitimacyReported2019, chenCryptocurrencyExchangesFake2022}, wash trading \cite{congCryptoWashTrading2019, pennecWashTradingCryptocurrency2021}, market manipulation \cite{gandalPriceManipulationBitcoin2018, chenMarketManipulationBitcoin2019, petersonMoonHistoryBitcoin2020}, pump-and-dump schemes,\cite{liCryptocurrencyPumpandDumpSchemes} \textit{etc}., although we should not lose sight of the fact that these ill-conceived practices, including the terminology, have their origins in traditional markets \cite{jarrowMarketManipulationBubbles1992, bradshawPumpDumpEmpirical2003, kimWhyPriceLimits2010}. That is to say, in our view, open ledgers, data, connectivity and trading accessible to everyone is a model that is much more likely to weed out bad actors more quickly and effectively, leading to free and open markets with efficient price discovery. The alternative would be the current mode of operation, where access to markets is restricted and obscured behind several intermediaries resulting in rent-seeking behavior with price discovery taking a back seat.\cite{khwajaRentSeekingCorruption2011}

\section{Open Interest and Perpetual Contracts}
\label{sec:oi_and_perps}

Open interest is the total number of outstanding contracts in a market. Similarly to trading volume it is an indicator of activity in that market\footnote{The definition of volume used in this study is the standard definition of volume reported by exchanges for
every trade. We refer the reader not familiar with this notion to the following definition \href{https://academy.binance.com/en/glossary/volume}{Binance/Glossary/Volume}.}. By definition, open interest can only change when the number of outstanding contracts changes. This can only happen in the following situations:

\begin{itemize}
    \item{} During settlement of futures contracts, when all open interest for that contract goes to zero.   
    \item{} During open interest increases when new contracts are created, \textit{i.e.}, by a pair of counterparties entering in a contract, one long and the other short.
    \item{} During decreases, when a party that holds a position trades in a way that reduces their exposure with another party that holds the opposite position. For example, if Trader A that has a $\$100$ long position in a contract reduces his exposure by say $\$50$ with another Trader B who has a $\$50$ short position, the open interest in the aforementioned contract will reduce by $\$50$ in value.
\end{itemize}

Open interest in perpetuals is computed the same way as for futures contracts with the distinction that, as these contracts do not expire, any increase or decrease of open interest in perpetual contracts implies that a prior trade has taken place. To better understand open interest in this context, it would be useful if we enumerate the possible interactions of market participants and their impact to open interest. In what follows, by "market" we mean a (trading pair, exchange) tuple, \textit{e.g.}, BTC/USDT linear perpetual on Binance or (\btcusdt{}, Binance). Consider the following transactions:

\begin{enumerate}[label=\alph*)]
    \item Trader $i$, who has no open positions in a market, opens a long position of $\$100$ on (\btcusdt{}, Binance) with trader $j$ who takes the other side of the trade, i.e., short. Given that neither trader $i$ nor $j$ had any previous positions this transaction increases the open interest, in value, by $\$100$.
    \item Trader $i$, who holds the above long position, now decides to reduce his exposure by $\$40$. He trades with a trader $k$ who holds no position in the (\btcusdt{}, Binance) market. Namely, trader $k$ goes long $\$40$ and trader $i$ short $\$40$, leaving him with a post-transaction net position of $\$60$ long (\btcusdt{}, Binance). In this case, the open interest remains unchanged as no contracts where created or settled. This transaction can be seen as a transfer of a $\$40$ long from trader $i$ to trader $k$.
    \item Now let us assume that trader $i$ would like to reduce his position by $\$20$. This time, however, he happens to be matched with trader $l$ who has a short position on (\btcusdt{}, Binance) of $\$40$ and would like to reduce it by $\$20$. In this interaction, open interest decreases by $\$20$ as this amount of contracts is effectively settled.
\end{enumerate}

There are only $3$ possibilities on every transaction (trade): a) increase, b) no change, or c) reduction in open interest. This can be described more formally with the following equation relating trading volume (transactions) and changes in open interest:

\begin{equation}
    \sum_{k>t_{i}}^{t_{i+1}} V_k \geq |O_{t_{i+1}} - O_{t_{i}}|
    \label{eq:oi_change}
\end{equation}
where $V_k$ is the size of a trade at time $t_k \in (t_i, t_{i+1}]$ and $O_{t}$ is the open interest at time $t$. \eqref{eq:oi_change} is summarizing what was described above, \textit{i.e.}, assuming that at time $t_{i}$ the open interest reported for a market is $O_{t_{i}}$ then the absolute change in open interest $|O_{t_{i+1}} - O_{t_{i}}|$ cannot be larger than the total trading volume $\sum_{k>t_{i}}^{t_{i+1}} V_k$ in that time interval $(t_i, t_{i+1}]$. To illustrate the mechanisms behind (\ref{eq:oi_change}), let us consider our previous examples and assume that no other trade took place between $t_{i}$ and $t_{i+1}$. For these transactions we have

\begin{enumerate}[label=\alph*)]
    \item{} $V_k = O_{t_{i+1}} - O_{t_{i}} = \$100$;
    \item{} $V_k = \$40$ and $O_{t_{i+1}} - O_{t_{i}} = \$0$;
    \item{} $V_k = -(O_{t_{i+1}} - O_{t_{i}}) = \$20$.
\end{enumerate}

From the previous examples, we can see that in all cases $V_k \geq |O_{t_{i+1}} - O_{t_i}|$. Equation (\ref{eq:oi_change}) is a consequence of taking the sum on both sides of the inequality and applying the triangle inequality on the right side when several transactions take place between $t_{i}$ and $t_{i+1}$. If the reporting of open interest follows the sequence of events, \textit{i.e.}, after a trade or liquidation the open interest is updated using the same timestamp as the event that effected the change, the above equation must hold for all time intervals\footnote{A liquidation is a forced trade by the exchange which takes place when one of the counter-parties has reached critical margin levels and can no longer maintain the position open. Liquidations are an automated risk reduction mechanism.}. Namely, every absolute change in open interest must be less than or equal to the trading volume of that time interval. However, as there is no single reporting standard in cryptocurrency exchanges, it is possible that in some intervals part of the trading volume responsible for the change in open interest is reported in the next time interval. For example, let us say that an exchange has two different API (Application Programming Interface) endpoints for reporting trades and the open interest for a market, let us further assume that it timestamps the API messages at the time of transmission and not as they occur. If the exchange receives large trades near $t_{i+1}$ and pushes the new open interest first and then, after that, the trades that caused the change, the trading volume in the time interval $(t_i, t_{i+1}]$ could be less than the change in open interest. However, this delay should be no more than a few multiples of the risk and matching engine latency, which for cryptocurrency exchanges is on the order of $500\mu s$\footnote{The matching and risk engines are crucial components of an exchange platform. The first one ensures that trades are executed efficiently and fairly following the principles of price-time priority. The second one ensures the safety and security of all funds and counter-party solvency by triggering liquidations when needed and ensuring that new positions do not exceed available margin.}. Given that the timestamp resolution for tick-by-tick data in the exchanges we investigate in this work is $1ms$, the delay should be no more than $1ms$ on reported trades (assuming honest reporting of volume). Even if the websocket API pushes data at a higher rate, which it does, the time resolution is the same. Since the exchanges presented in this study do not offer finer granularity timestamps, at least not according to their API documentation (see \textit{e.g.}, \href{https://binance-docs.github.io/apidocs/spot/en/#general-info}{Binance API/General Info}), we cannot know with certainty where the discrepancies presented in Section \ref{sec:analysis} are coming from: it can be misreporting of the traded volume (liquidations being part of the trading volume), the open interest, or both.

In this study we consider fixed time intervals $\delta$ over a given period $P$ and study changes in open interest in relation to traded volume. The first period is January of 2023 and the second one is July to September of 2023 (inclusive). For all $i$, $\delta \triangleq t_{i+1} - t_i$ is constant and the period $P$ and $\delta$ are related by $P = \delta N$. In what follows we consider the following choices for $\delta$ and $P$ ($N$ being determined by those two parameters)

\begin{itemize}
    \item{} $P = \delta = 1 \, \text{month}$, and therefore $N = 1$ as we consider the entire first period (January 2023 -- see Table \ref{tbl:oi_delta_period_1});
    \item{} $P = \delta = 3 \, \text{months}$ for the entire second period (July - September 2023 -- see Table \ref{tbl:oi_delta_period_2});
    \item{} $\delta = 1D$ (one day), $\delta = 1H$ (one hour) and finally $\delta = 1min$ (one minute) inside the first and second period -- see Table \ref{tbl:oi_delta_subperiod_1} and \ref{tbl:oi_delta_subperiod_2} respectively.
\end{itemize}

Another useful definition is that of \textit{minimal trading volume}. We define it as follows:

\begin{equation}
    mTV_{t_{i+1}} \triangleq |O_{t_{i+1}} - O_{t_{i}}|
    \label{eq:mtv}
\end{equation}

$mTV_{t_{i+1}}$ is the minimal trading volume required to have an observed absolute change in open interest equal to $|O_{t_{i+1}} - O_{t_{i}}|$ in the time interval $(t_i, t_{i+1}]$. Extending this to longer time intervals (\textit{e.g.}, 1 day, 1 hour, \textit{etc.}) we get the \textit{open interest total variation}
\begin{equation}
    O_{TV}^{P} \triangleq \sum_{i=1}^{N} |O_{t_{i+1}} - O_{t_{i}}| = \sum_{i=1}^{N} mTV_{t_{i+1}} \triangleq mTV^{P}
    \label{eq:oi:total_variation}
\end{equation}
which is also equal to the cumulative minimal trading volume, $mTV^{P}$. This latter quantity is a lower bound on the total trading volume in a time interval $P$ required to produce an observed open interest total variation equal to $O_{TV}^{P}$. For completeness, we define the \textit{total trading volume} for a time interval $P$ as:
\begin{equation}
    V_T^{P} \triangleq \sum_{i=1}^N\sum_{k=t_{i}}^{t_{i+1}} V_k
    \label{eq:total_trading_volume}
\end{equation}
Again this includes regular trades, block trades as well as liquidations. Equipped with these definitions and \eqref{eq:oi_change}, it follows that
\begin{equation}
    V_T^P \geq mTV^{P}
    \label{eq:volume_mtv}
\end{equation}

This is to say that the observed trading volume for a time period $P$ must be greater than or equal to the minimal trading volume required to produce the observed open interest total variation in the same period. If \eqref{eq:volume_mtv} does not hold at given time period $P$, this can be due to: i) not all trading volume being reported, ii) the open interest being incorrect, or iii) some combination of i) and ii). Arguably, hiding trading volume is not in the best interest of exchanges because they compete for liquidity, and trading volume is the single most important measure that market participants consider when gauging the liquidity of an exchange. Although this argument applies to regular trading volume, the same cannot necessarily be claimed for trading volume that has an associated conflict of interest, \textit{e.g.}, block trades by big clients or liquidations that could be seen, if frequent, as embarrassing to the exchange.

To aid us in the detection of any such violations we additionally define the \textit{excess open interest total variation} for the time period $P$ as:
\begin{equation}
    X_{TV}^{P} \triangleq \max\left\{ \sum_{i=1}^{N} |O_{t_{i+1}} - O_{t_{i}}| - \sum_{i=1}^N\sum_{k=t_{i}}^{t_{i+1}} V_k, 0\right\} = \max\{ O_{TV} - V_{T} , 0\}
    \label{eq:oi:excess_total_variation}
\end{equation}

When $X_{TV}^{P} > 0$, this means that for period $P$ the cumulative absolute changes in open interest cannot be explained by the reported trading volume. Said differently, assuming open interest is correct, then $X_{TV}^{P}$ is a lower bound on the \textit{missing} trading volume given the observed changes in open interest. Alternatively, if the observed volume is assumed to be correct, then $X_{TV}^P$ is the cumulative overstatement of the absolute open interest changes.

\section{Data}
\label{sec:data}

\begin{table}[!h]
\begin{center}
\caption{\label{tbl:markets}List of markets considered in this work.}

\begin{tabular}{lll p{10.0cm}}
\toprule
Exchange & Symbol & Contract Kind & Symbol URL \\
\midrule
\href{https://www.bybit.com/en-US/}{ByBit} & BTC\_USD\_IP & Inverse Perpetual & {\footnotesize \url{https://www.bybit.com/trade/inverse/BTCUSD}} \\
\href{https://www.bybit.com/en-US/}{ByBit} & BTC\_USDT\_P & Linear Perpetual & {\footnotesize \url{https://www.bybit.com/trade/usdt/BTCUSDT}} \\
\href{https://www.deribit.com/}{Deribit} & BTC\_USD\_IP & Inverse Perpetual & {\footnotesize \url{https://www.deribit.com/futures/BTC-PERPETUAL}} \\
\href{https://www.binance.com/en}{Binance} & BTC\_USD\_IP & Inverse Perpetual & {\footnotesize \url{https://www.binance.com/en/delivery/BTCUSD_PERPETUAL}} \\
\href{https://www.binance.com/en}{Binance} & BTC\_USDT\_P & Linear Perpetual & {\footnotesize \url{https://www.binance.com/en/futures/BTCUSDT}} \\
\href{https://www.bitmex.com/}{BitMEX} & BTC\_USD\_IP & Inverse Perpetual & {\footnotesize \url{https://www.bitmex.com/app/trade/XBTUSD}} \\
\href{https://www.okx.com/}{OKX} & BTC\_USD\_IP & Inverse Perpetual & {\footnotesize \url{https://www.okx.com/trade-swap/btc-usd-swap}} \\
\href{https://www.okx.com/}{OKX} & BTC\_USDT\_P & Linear Perpetual & {\footnotesize \url{https://www.okx.com/trade-swap/btc-usdt-swap}} \\
\href{https://pro.kraken.com/}{Kraken} & BTC\_USD\_IP & Inverse Perpetual & {\footnotesize \url{https://futures.kraken.com/trade/futures/PI_XBTUSD}} \\
\href{https://pro.kraken.com/}{Kraken} & BTC\_USD\_P & Linear Perpetual & {\footnotesize \url{https://futures.kraken.com/trade/futures/PF_XBTUSD}} \\
\href{https://www.huobi.com/en-us/}{HTX} & BTC\_USD\_IP & Inverse Perpetual & {\footnotesize \url{https://futures.htx.com/en-us/futures/swap/exchange#symbol=BTC}} \\
\href{https://www.huobi.com/en-us/}{HTX} & BTC\_USDT\_P & Linear Perpetual & {\footnotesize \url{https://futures.htx.com/en-us/futures/linear_swap/exchange#contract_code=BTC-USDT&contract_type=swap&type=cross}} \\
\bottomrule
\vspace{2em}
\end{tabular}
\end{center}
\end{table}

Our datasets are comprised of tick-by-tick trades, block trades, liquidations, and open interest as reported by the APIs of the respective exchanges mentioned in \tblref{tbl:markets}. We limit our attention to Bitcoin linear perpetuals quoted in \href{https://tether.to/en/}{USDT} ({\url{https://tether.to/en/})} and inverse perpetuals quoted in USD, as these are the most liquid derivatives. We focus on two periods: i) 2023/01/01 to 2023/01/31 (period 1) which is the beginning of the year and is usually a period of naturally higher trading volume and ii) 2023/07/01 to 2023/09/30 (period 2) containing most of the summer of months of $2023$ and September as it is the most recent month prior to this work which enables us to see if our observations are still pertinent in recent data. The infrastructure as well as the collected data are proprietary; however, in the interest of encouraging reproduction of this work, we offer a few suggestions on free and open source resources that can help in that respect (please see the Appendix for further details).

The exchanges in \tblref{tbl:markets} have been selected on the basis of volume (as seen, for one example, at CoinMarketCap)\cite{CryptocurrencyPricesCharts}, the fact that they have been in operation relatively long enough, and that the trading community considers them reasonably legitimate venues for trading---notwithstanding any allegations by regulators that have yet to be proven \cite{SECvBinance2023}.

\section{Analysis}
\label{sec:analysis}

The first set of results can be seen in \tblref{tbl:oi_delta_period_1} and \tblref{tbl:oi_delta_period_2} for the periods 1 and 2 respectively. In these tables we show the open interest total variation as defined by \eqref{eq:oi:total_variation}, the total volume (\eqref{eq:total_trading_volume}) and the excess total variation (\eqref{eq:oi:excess_total_variation}) for the entire periods, namely these are cumulative results over the periods of January 2023 and July to (and including) September 2023 (periods 1 and 2, respectively). As can be seen for period 1, only \bybit{} exhibits excess total variation on this scale while the rest of the exchanges seem to be reporting open interest changes that can be explained by the reported trading volume. In period 2 however, the results are significantly different (see \tblref{tbl:oi_delta_period_2}). In this period, both \bybit{} and OKX have non-negligible excess total variation. The excess in the \btcusdt{} and \btcusd{} markets on \bybit{} has grown quite dramatically, both in absolute terms and in proportion to the reported volume. Furthermore, in this period \btcusd{} on Binance starts showing signs of misreporting. The rest of the exchanges do not seem to have excess in this period either, at least when considering the period in its entirety. 

To put the observed excess on \bybit{} into perspective, if the open interest reported by it in period 2 is to believed for the \btcusdt{} market, the lower bound on the trading volume required to produce the observed changes in open interest would be more than $\$128bn$. However, this minimum requires that all trading volume in every reporting interval is in the direction that increases (or decreases) open interest. With so much volume, and presumably participants, it would be unrealistic to assume that level of trading synchronicity. If we apply ratios of $O_{TV}/V_T$ from exchanges that do not misreport open interest (at least not to that degree) and apply them to the $O_{TV}$ for the \btcusdt{} market on \bybit{}, that would imply trading volumes in the range of $\$156bn$ to $\$213bn$, \textit{i.e.}, greater than the volume on the \btcusdt{} market on Binance(!); which in our view is highly improbable.

\begin{table}[!h]
\begin{center}
\caption{\label{tbl:oi_delta_period_1}Open interest total variation, total trading volume and excess total variation for the period 2023/01/01 to 2023/01/31 (inclusive). For comparison purposes we have converted bitcoin (\bitcoin) units to USD using the average price of the exchange rate for the period which was $\$20,\mkern-3mu625$ per bitcoin.}
\resizebox{\textwidth}{!}{%

\begin{tabular}{lllll}
\toprule
Exchange & Symbol & $O_{TV}$ & $V_T$ & $X_{TV}$ \\
\midrule
ByBit & BTC\_USDT\_P & \bitcoin2,213,583 (\$45.66B) & \bitcoin1,469,962 (\$30.32B) & \bitcoin743,622 (\$15.34B) \\
ByBit & BTC\_USD\_IP & \$12,088,654,910 & \$6,570,819,230 & \$5,517,835,680 \\
Binance & BTC\_USDT\_P & \bitcoin2,084,275 (\$42.99B) & \bitcoin4,050,268 (\$83.54B) & \bitcoin0 (\$0) \\
OKX & BTC\_USDT\_P & \bitcoin905,525 (\$18.68B) & \bitcoin949,431 (\$19.58B) & \bitcoin0 (\$0) \\
BitMEX & BTC\_USD\_IP & \$4,065,203,600 & \$5,599,999,100 & \$0 \\
OKX & BTC\_USD\_IP & \$3,924,367,800 & \$5,325,455,400 & \$0 \\
Deribit & BTC\_USD\_IP & \$3,665,856,750 & \$4,045,272,670 & \$0 \\
HTX & BTC\_USDT\_P & \bitcoin133,615 (\$2.76B) & \bitcoin381,464 (\$7.87B) & \bitcoin0 (\$0) \\
Kraken & BTC\_USD\_P & \bitcoin25,895 (\$534.08M) & \bitcoin35,024 (\$722.37M) & \bitcoin0 (\$0) \\
HTX & BTC\_USD\_IP & \$234,509,100 & \$632,036,900 & \$0 \\
Kraken & BTC\_USD\_IP & \$201,790,503 & \$315,671,226 & \$0 \\
Binance & BTC\_USD\_IP & \$86,462,494 & \$120,899,920 & \$0 \\
\bottomrule
\end{tabular}}

\end{center}

\end{table}

\begin{table}[!h]
\begin{center}
\caption{\label{tbl:oi_delta_period_2}Open interest total variation, total trading volume, and excess total variation for the period 2023/07/01 to 2023/09/30 (inclusive). Similarly to the previous table we use the average price for the period to facilitate comparisons. The average price for the period was $\$28,\mkern-3mu250$ per bitcoin.}
\resizebox{\textwidth}{!}{%
\begin{tabular}{lllll}
\toprule
Exchange & Symbol & $O_{TV}$ & $V_T$ & $X_{TV}$ \\
\midrule
ByBit & BTC\_USDT\_P & \bitcoin4,583,448 (\$129.48B) & \bitcoin2,571,288 (\$72.64B) & \bitcoin2,012,160 (\$56.84B) \\
OKX & BTC\_USDT\_P & \bitcoin3,213,509 (\$90.78B) & \bitcoin2,598,848 (\$73.42B) & \bitcoin614,661 (\$17.36B) \\
ByBit & BTC\_USD\_IP & \$22,856,881,902 & \$10,417,205,910 & \$12,439,675,992 \\
OKX & BTC\_USD\_IP & \$11,568,080,200 & \$10,203,157,300 & \$1,364,922,900 \\
Binance & BTC\_USD\_IP & \$323,617,912 & \$285,387,614 & \$38,230,298 \\
Binance & BTC\_USDT\_P & \bitcoin5,899,253 (\$166.65B) & \bitcoin7,106,811 (\$200.77B) & \bitcoin0 (\$0) \\
BitMEX & BTC\_USD\_IP & \$12,067,917,700 & \$16,241,861,300 & \$0 \\
Deribit & BTC\_USD\_IP & \$10,316,342,380 & \$10,419,942,140 & \$0 \\
HTX & BTC\_USDT\_P & \bitcoin266,235 (\$7.52B) & \bitcoin741,751 (\$20.95B) & \bitcoin0 (\$0) \\
Kraken & BTC\_USD\_P & \bitcoin77,984 (\$2.2B) & \bitcoin107,353 (\$3.03B) & \bitcoin0 (\$0) \\
HTX & BTC\_USD\_IP & \$456,969,000 & \$1,289,953,500 & \$0 \\
Kraken & BTC\_USD\_IP & \$446,044,392 & \$733,864,191 & \$0 \\
\bottomrule
\vspace{1em}
\end{tabular}}

\end{center}
\end{table}

Considering that the periods are quite lengthy in comparison with the open interest reporting period in all the exchanges under consideration, we elected to refine the resolution in the hope of gleaning more insight in the observed excess total variation in Tables~\ref{tbl:oi_delta_period_1} and \ref{tbl:oi_delta_period_2}. With that objective in mind we chose three more sub-periods: i) one day ($1D$), ii) one hour ($1H$), and iii) one minute ($1min$), within periods 1 and 2. For each one of those sub-periods we computed: i) the probability of observing excess total variation for the sub-period $SP$ within each period $\mathbb{P}_{SP}(X_{TV} > 0)$, and ii) the expectation of the excess total variation conditional on this excess being greater than $0$: $\mathbb{E}_{SP}[X_{TV} | X_{TV} > 0]$. These results can be seen in \tblref{tbl:oi_delta_subperiod_1} and \tblref{tbl:oi_delta_subperiod_2} for periods 1 and 2 respectively. At these finer resolutions we see that virtually all the exchanges have less than ideal open interest reporting practices. For instance, we can see on \bybit{} that the reported open interest cannot be reconciled with the trading volume on any of the selected sub-periods. In essence, the open interest is incorrect every day, almost every hour, and in more than $70\%$ of the one minute sub-periods. It is particularly impressive that the expected excess total variation in the one minute sub-period for the \btcusdt{} market on \bybit{} has a size that is more than $\$500,\mkern-3mu000$. In fact, it is quite surprising that only HTX and Kraken (and perhaps to some degree \bitmex{}) report open interest that can almost be reconciled on every sub-period. 

\begin{table}[!h]
\begin{center}
\caption{\label{tbl:oi_delta_subperiod_1}Probability of excess total variation and expected excess total variation for the period 2023/01/01 to 2023/01/31 (inclusive). Average price $\$20,\mkern-3mu625$ per bitcoin.}
\resizebox{\textwidth}{!}{%
\begin{tabular}{llllllll}
\toprule
Exchange & Symbol & $\mathbb{P}_{1D}(X_{TV} > 0)$ & $\mathbb{E}_{1D}[X_{TV} | X_{TV} > 0]$ & $\mathbb{P}_{1H}(X_{TV} > 0)$ & $\mathbb{E}_{1H}[X_{TV} | X_{TV} > 0]$ & $\mathbb{P}_{1min}(X_{TV} > 0)$ & $\mathbb{E}_{1min}[X_{TV} | X_{TV} > 0]$ \\
\midrule
ByBit & BTC\_USD\_IP & 100.0\% & \$177,994,699  & 98.9\% & \$7,633,055  & 70.8\% & \$200,682  \\
ByBit & BTC\_USDT\_P & 100.0\% & \bitcoin23,988 (\$494.75M) & 98.5\% & \bitcoin1,043 (\$21.51M) & 72.4\% & \bitcoin28 (\$577.14K) \\
OKX & BTC\_USDT\_P & 51.6\% & \bitcoin1,992 (\$41.08M) & 70.8\% & \bitcoin159 (\$3.29M) & 51.2\% & \bitcoin9 (\$181.63K) \\
OKX & BTC\_USD\_IP & 12.9\% & \$6,585,500  & 31.3\% & \$805,784  & 29.9\% & \$59,521  \\
Deribit & BTC\_USD\_IP & 12.9\% & \$4,856,062  & 29.8\% & \$719,027  & 28.9\% & \$46,847  \\
Binance & BTC\_USD\_IP & 6.5\% & \$205,770  & 45.4\% & \$15,317  & 39.9\% & \$1,102  \\
Kraken & BTC\_USD\_P & 6.5\% & \bitcoin93 (\$1.92M) & 22.6\% & \bitcoin8 (\$163.46K) & 14.7\% & \bitcoin1 (\$16.56K) \\
Kraken & BTC\_USD\_IP & 3.2\% & \$166,194  & 12.0\% & \$39,292  & 8.7\% & \$5,996  \\
BitMEX & BTC\_USD\_IP & 0.0\% & \$0  & 17.2\% & \$702,462  & 28.3\% & \$57,394  \\
HTX & BTC\_USD\_IP & 0.0\% & \$0  & 1.5\% & \$54,136  & 6.2\% & \$9,267  \\
Binance & BTC\_USDT\_P & 0.0\% & \bitcoin0 (\$0) & 0.4\% & \bitcoin259 (\$5.34M) & 14.0\% & \bitcoin19 (\$398.15K) \\
HTX & BTC\_USDT\_P & 0.0\% & \bitcoin0 (\$0) & 0.0\% & \bitcoin0 (\$0) & 11.5\% & \bitcoin2 (\$35.8K) \\
\bottomrule

\end{tabular}}

\end{center}
\end{table}

\begin{table}[!h]
\begin{center}
\caption{\label{tbl:oi_delta_subperiod_2}Probability of excess total variation and expected excess total variation for the period 2023/07/01 to 2023/09/30 (inclusive). Average price $\$28,\mkern-3mu250$ per bitcoin.}
\resizebox{\textwidth}{!}{%
\begin{tabular}{llllllll}
\toprule
Exchange & Symbol & $\mathbb{P}_{1D}(X_{TV} > 0)$ & $\mathbb{E}_{1D}[X_{TV} | X_{TV} > 0]$ & $\mathbb{P}_{1H}(X_{TV} > 0)$ & $\mathbb{E}_{1H}[X_{TV} | X_{TV} > 0]$ & $\mathbb{P}_{1min}(X_{TV} > 0)$ & $\mathbb{E}_{1min}[X_{TV} | X_{TV} > 0]$ \\
\midrule
ByBit & BTC\_USDT\_P & 100.0\% & \bitcoin21,871 (\$617.86M) & 99.8\% & \bitcoin918 (\$25.92M) & 75.8\% & \bitcoin22 (\$627.06K) \\
ByBit & BTC\_USD\_IP & 100.0\% & \$135,213,869  & 99.6\% & \$5,658,882  & 70.1\% & \$146,323  \\
OKX & BTC\_USDT\_P & 96.7\% & \bitcoin7,070 (\$199.73M) & 92.7\% & \bitcoin343 (\$9.69M) & 59.7\% & \bitcoin13 (\$361.45K) \\
Binance & BTC\_USD\_IP & 88.0\% & \$579,256  & 83.2\% & \$37,878  & 48.7\% & \$2,232  \\
OKX & BTC\_USD\_IP & 85.9\% & \$21,183,446  & 72.5\% & \$1,395,272  & 42.2\% & \$72,655  \\
Deribit & BTC\_USD\_IP & 52.2\% & \$6,708,834  & 44.0\% & \$893,163  & 29.1\% & \$59,624  \\
Binance & BTC\_USDT\_P & 27.2\% & \bitcoin4,893 (\$138.22M) & 41.1\% & \bitcoin445 (\$12.56M) & 33.4\% & \bitcoin39 (\$1.09M) \\
Kraken & BTC\_USD\_P & 4.3\% & \bitcoin80 (\$2.26M) & 20.6\% & \bitcoin9 (\$268.19K) & 15.6\% & \bitcoin1 (\$20.91K) \\
BitMEX & BTC\_USD\_IP & 3.3\% & \$5,323,233  & 28.5\% & \$579,009  & 32.0\% & \$54,281  \\
Kraken & BTC\_USD\_IP & 0.0\% & \$0  & 6.4\% & \$34,787  & 5.9\% & \$6,331  \\
HTX & BTC\_USD\_IP & 0.0\% & \$0  & 1.0\% & \$62,557  & 4.3\% & \$6,371  \\
HTX & BTC\_USDT\_P & 0.0\% & \bitcoin0 (\$0) & 0.4\% & \bitcoin19 (\$526.5K) & 11.3\% & \bitcoin2 (\$43.75K) \\
\bottomrule
\vspace{1em}
\end{tabular}}

\end{center}
\end{table}

We can also see that the expected excess total variation seems to be increasing in period 2 (P2) compared with period 1 (P1) on some exchanges. For example, on OKX for the \btcusdt{} market $\mathbb{E}_{1D}[X_{TV} | X_{TV} > 0]$ is $\sim 4$ times higher for the $1D$ sub-period and $\sim2$ times higher when considering the $1H$ sub-period. A similar pattern can be seen on \bybit{}, Binance, and to some extent on Deribit and HTX, although on Deribit it is considerably less pronounced. Namely, incongruities are amplified during the summer months on some exchanges, at least in the two periods considered in this work.

One analysis that we considered performing was to normalize the expectations in Tables~\ref{tbl:oi_delta_subperiod_1} and \ref{tbl:oi_delta_subperiod_2} by the mean trading volume typical for that sub-period; however, given that we cannot be sure if it is the open interest that is incorrect or the reported trading volume (or both), it is preferable that all quantities remain in absolute and not relative terms.

\begin{figure}
    \centering
    \includegraphics[width=0.7 \paperwidth, bb=0 0 1600 700]{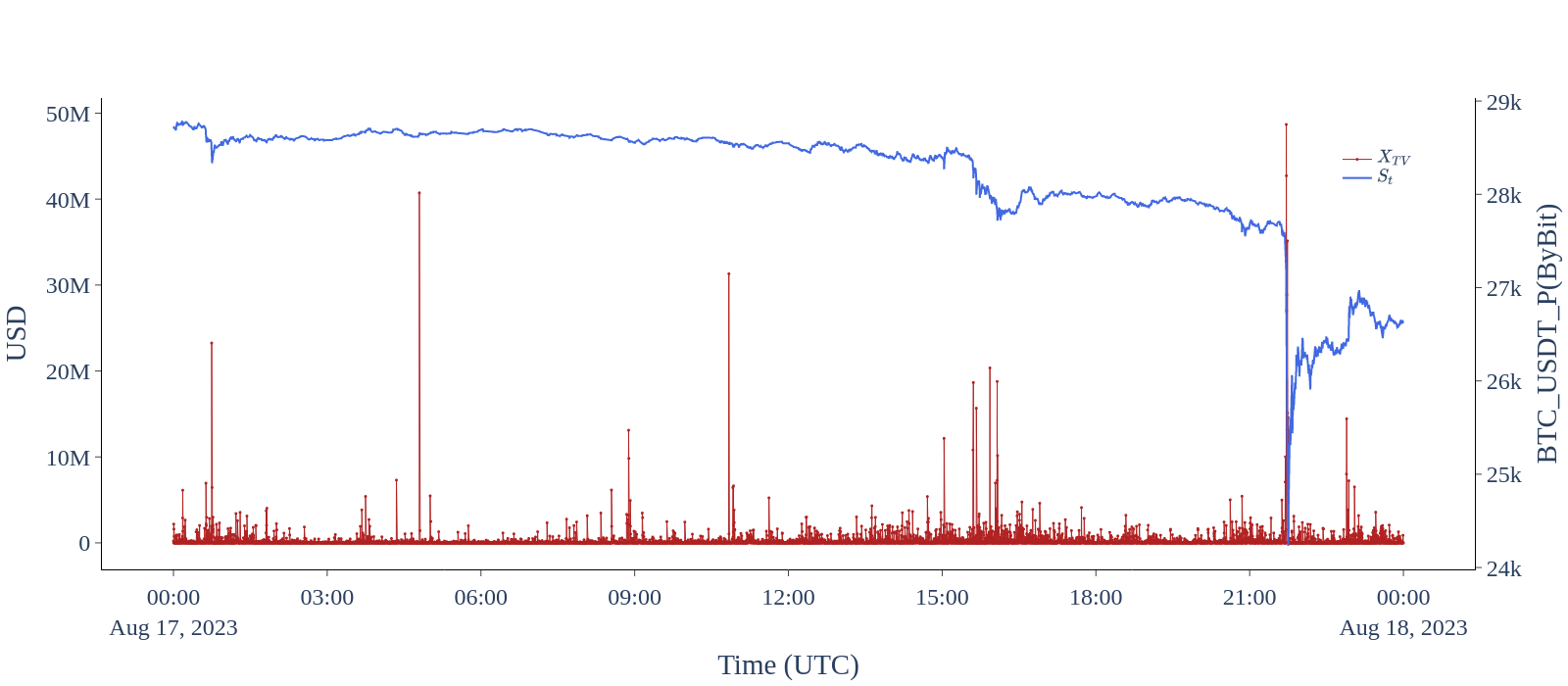}
    \caption{Excess open interest total variation ($X_{TV}$) computed on a tick-by-tick basis, namely on every open interest update for the \btcusdt{} market on \bybit{} for 2023/08/17.}
    \label{fig:bybit_aug_17}
\end{figure}

\begin{figure}
    \centering
    \includegraphics[width=0.7 \paperwidth, bb=0 0 1600 700]{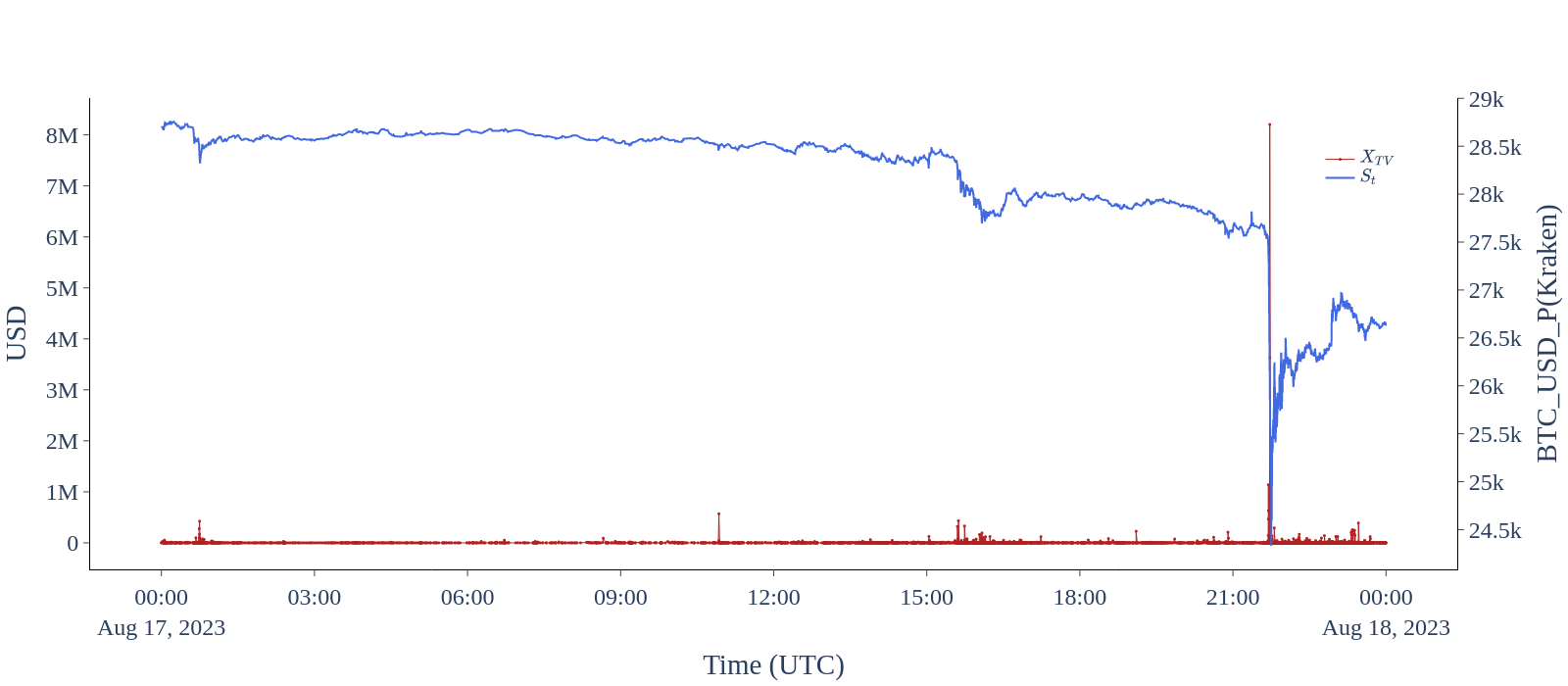}
    \caption{Excess open interest total variation ($X_{TV}$) computed on a tick-by-tick basis, namely on every open interest update for the \btcusdt{} market on Kraken for 2023/08/17.}
    \label{fig:kraken_aug_17}
\end{figure}

\section{Discussion}
\label{sec:discussion}

First let us summarize our findings from Section~\ref{sec:analysis}.

\begin{itemize}
    \item When we consider large periods, only \bybit{} exhibits irreconcilable discrepancies between open interest and trading volume in period 1. In period 2 this set of exchanges increases to \bybit{}, OKX and Binance.
    \item If we consider sub-periods within periods 1 and 2 we cannot reconcile open interest with trading volume on any exchange.
    \item Allowing some \textit{margin of error}, we can reconcile the above quantities only on HTX and Kraken.
    \item We see that some exchanges seem to be changing their behavior in different periods: for example, OKX, Deribit and Binance had fewer discrepancies in period 1 when compared with period 2 almost uniformly on all sub-periods.
    \item Exchanges that have an almost perfect record on longer periods start exhibiting discrepancies on lower sub-periods, \textit{e.g.}, HTX, Kraken and \bitmex{}.
    \item The magnitude of the conditional excess total variation in period 2 is relatively higher compared with period 1 on OKX, \bybit{}, Binance, Deribit and HTX.
\end{itemize}

The question that therefore presents itself is why open interest cannot be reconciled with trading volume. Although we cannot offer a definitive answer, since that would require insider information on the systems and policies of all the exchanges mentioned, we can recount the possibilities. As briefly mentioned in Section~\ref{sec:oi_and_perps}, either i) not all trading volume is reported, ii) the open interest is incorrect, or iii) some combination of the two is taking place.

Let us begin by considering the first case, namely that the reported trading volume is less than the actual trading volume on perpetual swaps, while open interest is reported accurately. This could be the case due to several reasons. One is if the exchanges have a delay of more than one minute in reporting some volume, \textit{e.g.}, block trades or liquidations, and perhaps even regular trades. If this is what is happening then we would expect to observe eventual consistency on higher timeframes, which is to say that the probability of observing excess total variation should become progressively lower from the $1min$ toward the full period. For some exchanges this is exactly what is happening, see for example the \btcusdt{} market on HTX in \tblref{tbl:oi_delta_subperiod_1} and \tblref{tbl:oi_delta_subperiod_2}. If, however, the missing volume is not delayed but in fact \textit{never} reported, this would lead to $\mathbb{P}_{SP}(X_{TV} > 0)$ increasing from lower timeframes to higher timeframes. This is consistent with what we observe on \btcusd{} and \btcusdt{} on \bybit{}, OKX, and Binance, and on \btcusd{} on Deribit. If this is true, namely some volume is never reported while the open interest is accurate, then the question is: why? On the surface it would appear to be contrary to the exchange's self-interest: indeed, the literature would suggest that exchanges attempt to do exactly the opposite \cite{congCryptoWashTrading2019, InvestigationLegitimacyReported2019, chenCryptocurrencyExchangesFake2022}. However, one possible explanation would be if the exchanges and/or affiliated entities are entering positions that they do not want to disclose to other market participants, \textit{e.g.}, partial reporting of block trades. This, as evidenced by the FTX case, would not be outside the realm of possibility.\cite{j.rayiiiCaseNo2211068} If that is indeed what is happening, then the amount of additional risk in the books of said entities would be staggering. Another possibility is that all or a portion of liquidations is never reported, as a high number of liquidations on an exchange could potentially alarm market participants, resulting in more cautious behavior, lower trading volume, and a subsequent reduction in revenue streams for the exchange.

The other possibility is that trading volume is accurate but the reported open interest is incorrect. Market participants do observe the changes in open interest and attempt to infer how \textit{informed} investors are being positioned in the market. The general heuristic is that if open interest is rising and the price is increasing then the aggressors, presumably informed investors, are the buyers. If on the other hand open interest is rising and the price is falling, the aggressors would be the sellers. If, however, as the price increases open interest decreases, this presumably implies shorts covering (and implies the same for the longs when price is dropping). This heuristic may or may not be valid, but its validity is less important than whether market participants pay attention to it and whether some trade according to it. Given that exchanges generate revenue by extracting fees on traded volume, it is conceivable that they could potentially generate \textit{false signals}, when the market has none to offer, by modulating open interest artificially, with the expectation that this would incentivize market participants to trade more. If that is what actually takes place, it would require those \textit{signals} to be as clear and large in magnitude as possible, such that all market participants notice them. In this case we would expect $\mathbb{E}_{SP}[X_{TV} | X_{TV} > 0]$ to be inflated, similar to what we see for \bybit{}, OKX, and Binance in \tblref{tbl:oi_delta_subperiod_1} and \tblref{tbl:oi_delta_subperiod_2} for all sub-periods. Although such activity would be almost impossible to prove beyond reasonable doubt it is instructive to look at the $X_{TV}$ on the tick-by-tick level during an eventful day, \textit{e.g.}, a significant market decline. On August 17, 2023 the Bitcoin price crashed more than $10\%$ during the US trading session. In Figure 1, we see this event as observed on \bybit{}, and in Figure 2 we see the same day as it unfolded on Kraken. In both plots the red lines represent the excess total variation observed, with the left axis measuring the magnitude of the excess in USD. The right axis represents the price, and the blue line is the last traded price at the time of the open interest update. Comparing the two figures it is immediately evident that on \bybit{} there seem to be almost no time intervals where $X_{TV} = 0$, while on Kraken this condition holds almost for the entirety of the day. The few spikes in $X_{TV}$ on Kraken could conceivably be explained by delayed reporting of liquidations or open interest, as most of them are localized in time intervals with sharp price fluctuations.

\section{Conclusions}

In this work we consider the most liquid Bitcoin perpetual swaps on seven of the top cryptocurrency exchanges. 
We find that trading volume cannot be reconciled with the reported changes in open interest for the majority of these exchanges. It is unclear whether this is due to delayed or unreported trading volume or due to incorrectly reported open interest. In our view, the most likely scenario is that both are true, perhaps, however, not to the same degree on every exchange. Although we could not perfectly reconcile these quantities for any of the exchanges in question, we find that there are discernible differences in behavior across these exchanges. The discrepancies on \bybit{} and OKX are so frequent and large in magnitude that these two exchanges merit a category of their own. On these exchanges we could not reconcile trading volume with reported open interest in any time period, with the implied trading volume being in the range of hundreds of billions over and above the reported trading volume, assuming the open interest is the quantity that is correct. If in fact, however, the trading volume is the more accurately reported quantity, this would imply that the open interest on these exchanges is almost completely fabricated. This could perhaps be explained by certain incentive structures baked into the scenario: leading market participants to believe that \textit{informed} investors are taking large positions in these markets (as implied by the large change in open interest) could---depending on the participants' prior positioning---lead to panic or fear of missing out on potential profits, thereby increasing trading volume, and profit for the exchange. Given that volatility and trading volumes in Bitcoin and other cryptocurrencies have been trending lower in $2023$ we believe that the latter is a more plausible explanation. Figures~\ref{fig:bybit_aug_17} and \ref{fig:kraken_aug_17} also seem to point in that direction.

Binance, Deribit and \bitmex{} form, conceptually, another cluster of exchanges. Although we could not reconcile the changes in open interest with trading volume, the frequency and magnitude of the discrepancies is such that it leaves room for some relatively more benign explanation (see Section~\ref{sec:discussion}). The last group of exchanges is formed by Kraken and HTX, who have the lowest number of discrepancies. For these exchanges we could reconcile changes in open interest with trading volume on almost all sub-periods (see Tables~\ref{tbl:oi_delta_subperiod_1} and \ref{tbl:oi_delta_subperiod_2}).

What applies to all exchanges, however, is that even if there exists some explanation, the fact remains that market data is systematically misreported to varying degrees---be it in the form of a delay, omission, or fabrication. Such inaccuracies could easily be exploited by these exchanges to manipulate the market participants' perception of price evolution, even if it is for short periods of time. If there is a lesson to be learned by the relatively recent demise of Alameda and FTX, it is that if something can be exploited, it most likely will be exploited.\cite{j.rayiiiCaseNo2211068} The solution to this would be, from a technical perspective, very easy to implement: simply add open interest to every trade pushed out by the market feeds, and report all trades without delay---irrespective of their kind---to all market participants with the correct price, size, and timestamp of when the trade took place, ideally with microsecond precision. As open interest is, in a way, a representation of the total outstanding liabilities in a market, it should be treated with the same rigor as proof of reserves.

Lastly, we would like to call upon the exchanges examined in this work to evaluate carefully the evidence presented herein, in the hope that they will take our suggestions into consideration and demonstrate with their actions their commitment to their users and free and open markets. If this industry is to grow and flourish, we would be well served to remember the principles and ethos of the people that put in place its foundation \cite{nakamotoBitcoinPeertoPeerElectronic2008}.

\appendix

\section{Open Source API Connectors and Data Sources}
\label{apdx:data}

There are many libraries and tick-by-tick data providers for the cryptocurrency markets. We cannot endorse any commercial products. We restrict our focus on free and open source libraries as well as data sources.

In terms of libraries we would recommend \href{https://github.com/ccxt/ccxt}{ccxt}\footnote{\url{https://github.com/ccxt/ccxt}} which has connectivity to more than $90$ cryptocurrency exchanges, including the exchanges that we focus on this work. This library should prove sufficient for collecting the data on which this work is based on. If a custom solution is required, as we have already mentioned, each of these exchanges has an API that users can collect data even without account. We refer our readers to the respective exchange websites that are available in \tblref{tbl:markets} for the REST and websocket API documentation.

Binance, which is the leader in this space has gone the extra mile and offers level 2 data directly on their website. Order book data has to be requested specifically, but as of the of this writing Binance does not apply any charge for such data. These are available at the \href{https://www.binance.com/en/landing/data}{Binance Historical Market Data}\footnote{\url{https://www.binance.com/en/landing/data}}.

\bibliographystyle{apalike}
\bibliography{bibliography}

\end{document}